\documentclass[fleqn,usenatbib]{mnras}

\usepackage[T1]{fontenc}
\usepackage{ae,aecompl}

\usepackage{amsmath}
\usepackage{graphicx}
\usepackage{amssymb}

\title{The Euclidean distribution of Fast Radio Bursts}

\author[N. Oppermann et al.]{Niels Oppermann,$^{1,2}$\thanks{E-mail: niels@cita.utoronto.ca (NO)}
Liam D.~Connor,$^{1,2,3}$ Ue-Li Pen$^{1,2,4,5}$
\\
$^{1}$Canadian Institute for Theoretical Astrophysics, University of Toronto, 60 St.\ George Street, Toronto ON, M5S 3H8, Canada\\
$^{2}$Dunlap Institute for Astronomy and Astrophysics, University of Toronto, 50 St.\ George Street, Toronto ON, M5S 3H4, Canada\\
$^{3}$Department of Astronomy and Astrophysics, University of Toronto, 50 St.\ George Street, Toronto ON, M5S 3H4, Canada\\
$^{4}$Canadian Institute for Advanced Research, 180 Dundas St.\ West, Toronto ON, M5G 1Z8, Canada\\
$^{5}$Perimeter Institute for Theoretical Physics, 31 Caroline St.\ North, Waterloo ON, N2L 2Y5, Canada
}

\date{Accepted XXX. Received YYY; in orignal form ZZZ}

\pubyear{2016}

\begin{document}
\label{firstpage}
\pagerange{\pageref{firstpage}--\pageref{lastpage}}
\maketitle

\begin{abstract}
	We investigate whether current data on the distribution of observed flux densities of Fast Radio Bursts (FRBs) are consistent with a constant source density in Euclidean space. We use the number of FRBs detected in two surveys with different characteristics along with the observed signal-to-noise ratios of the detected FRBs in a formalism similar to a $V/V_\mathrm{max}$-test to constrain the distribution of flux densities. We find consistency between the data and a Euclidean distribution. Any extension of this model is therefore not data-driven and needs to be motivated separately. As a byproduct we also obtain new improved limits for the FRB rate at 1.4\,GHz, which had not been constrained in this way before.
\end{abstract}


\section{Introduction}

Fast Radio Bursts (FRBs) are bright, short radio pulses that are highly dispersed. Most FRBs observed so far have dispersion measures that are much larger than what is expected due to the Milky Way's interstellar medium. Thus, it is likely that the sources producing these bursts are extragalactic. If most of the dispersion is due to the ionised intergalactic medium, the sources have to be at redshifts of the order of one \citep[e.g.,][]{champion-2015}. In this case, they may soon become useful as cosmological probes \citep[e.g.,][]{masui-2015a}. However, the large observed dispersion may also be due to dense material surrounding the source \citep[e.g.,][]{connor-2016a}, putting them at potentially much smaller distances.

An observable distinction between cosmological models and more local models for the FRB population is given by the distribution of flux densities. This is independent of the pulse dispersion. The simplest model is one in which FRBs are uniformly distributed and their location does not influence any of their observable properties other than that farther bursts appear dimmer. In Euclidean space this leads to an unambiguous prediction for their flux densities, namely
\begin{equation}
	\mathrm{d}N \propto S^{-(\alpha + 1)} \, \mathrm{d}S
	\label{eq:powerlaw}
\end{equation}
with $\alpha = 3/2$, where $S$ is the flux density and $\mathrm{d}N$ is the number of FRBs with a flux density in an infinitesimal interval $\mathrm{d}S$. This relation holds independently from their intrinsic luminosity distribution and does not require them to be standard candles.
Deviations from an index 3/2 are for example expected if the number density of FRBs is not constant and/or if their distances are large enough that deviations from Euclidean geometry become important. If FRBs are extragalactic, but local, then the default expectation should be $\alpha = 3/2$. It is worth noting that a constant-density Euclidean distribution is the only model that leads to a clear prediction for the observed flux densities. If FRBs are at large enough distances that the expansion history of the Universe is important, then the population of FRBs can be expected to also have undergone some evolution, thus introducing functional degrees of freedom that are at present completely unconstrained. An additional effect that can in principle influence the measured distribution of flux densities is dispersion smearing, i.e., the dispersion of the pulses' arrival times within a frequency bin of a survey, resulting in a reduction of the signal-to-noise ratio for very high dispersion measures. If dispersion measures were correlated with distance, this could potentially lead to a systematic flattening of the distribution, but will not affect its shape if the dispersion measure and distance of an FRB are uncorrelated.

In this note we investigate whether the constant-density Euclidean model is consistent with current observations. We do this by studying the one-parametric class of models in which $\alpha$ is a free parameter and which includes the constant-density Euclidean model as a special case. Deriving the observational constraints on the parameter $\alpha$ has the potential of showing that $\alpha=3/2$ is disfavoured by the data without the need to make any more complicated model assumptions. We derive constraints from a combination of source counts of different surveys and the observed $V/V_\mathrm{max}$-values \citep{schmidt-1968}.

In the simple constant-density Euclidean model, the power-law behaviour of Eq.~\eqref{eq:powerlaw} is not only valid for the flux densities, but also for any other observable that depends linearly on flux density. One example would be the fluence
\begin{align}
	F &= \int\mathrm{d}t \, S(t)\nonumber\\
	&= S \, \tau,
\end{align}
where $\tau$ is the duration of a burst and $S$ is the average flux density within the interval of length $\tau$. Another example is the observed signal-to-noise ratio, which depends on the flux density of the burst, its duration, and of course on properties of the telescope and the survey. In many cases it can be approximated as
\begin{equation}
	s \approx K \, S \, \tau^{1/2},
\end{equation}
where we include all instrumental properties in the constant $K$ \citep[e.g.,][]{caleb-2016}. The detection of an FRB is always subject to a sensitivity cutoff in signal-to-noise, and not in flux density or fluence. This makes the statistics of flux density and fluence more complicated than the statistics of the signal-to-noise ratio $s$ and we choose to cast all equations in terms of $s$. Note that, as is common in the field, we use the term signal-to-noise ratio to mean the amplitude of the FRB signal plus noise, divided by the standard deviation of the noise, which can be approximately determined empirically for each search window length $\tau$.

We derive the necessary statistical methodology in the following section, discuss the data that we use and show our results in Sect.~\ref{sec:results}. We conclude with a discussion in Sect.~\ref{sec:discussion}.

\section{Methodology}
\label{sec:methods}

\subsection{Likelihood for the observed signal-to-noise ratios}
\label{sec:snr}

Clearly some information on the parameter $\alpha$ is contained in the distribution of observed signal-to-noise ratios. If we assume a value for $\alpha$ and suppose that a survey with a given signal-to-noise threshold $s_\mathrm{min}$ detects an FRB, then the likelihood for its signal-to-noise ratio to be $s$ is, according to Eq.~\eqref{eq:powerlaw},
\begin{equation}
	\mathcal{P}(s|s_\mathrm{min},\alpha) = \left\{\begin{array}{cc}\frac{\alpha}{s_\mathrm{min}} \, \left(\frac{s}{s_\mathrm{min}}\right)^{-(\alpha + 1)} & \textnormal{if } s \geq s_\mathrm{min}\\0&\textnormal{else}\end{array}\right..
\end{equation}
If the survey detects $n$ independent FRBs then the joint likelihood for their signal-to-noise ratios is simply the product of the individual ones.

For $N$ different surveys, each detecting $n_1,\dots, n_{N}$ FRBs, the situation is the same, except that we have to take into account that each survey has a different detection threshold $s_\mathrm{min}$. If we denote the observed signal-to-noise value of the $i$-th FRB in the $I$-th survey as $s_{I,i}$, the $(n_1+\cdots+n_N)$-dimensional vector of all these observed values as $\vec{s}$, the $N$-dimensional vector of all threshold signal-to-noise values as $\vec{s}_\mathrm{min}$, and the $N$-dimensional vector of the numbers of detections in each survey as $\vec{n}$, the complete likelihood becomes
\begin{equation}
	\mathcal{P}(\vec{s}|\vec{n},\vec{s}_\mathrm{min},\alpha) = \prod_{I=1}^N \prod_{i=1}^{n_I} \mathcal{P}(s_{I,i}|s_{\mathrm{min},I},\alpha).
\end{equation}
This is easily calculated and we will do so in Sect.~\ref{sec:results}. In the following we will refrain from mentioning $\vec{s}_\mathrm{min}$ explicitly in the notation of probabilities and imply that all survey properties are always fixed.

Note that the combination $(s/s_\mathrm{min})^{-\alpha}$ for $\alpha = 3/2$ corresponds to the ratio of the volume interior to the FRB and the volume in which this particular FRB could have been detected by the survey, $V/V_\mathrm{max}$, for a constant source density in three-dimensional Euclidean space. The likelihood we are using here to constrain $\alpha$ is thus closely related to the $V/V_\mathrm{max}$-test used in many contexts to check for deviations from a constant density for a source population \citep[e.g.,][]{schmidt-1968}.

\subsection{Likelihood for the number of observed FRBs}
\label{sec:number}

In addition to the information contained in the signal-to-noise ratios of the observed bursts, some information is also contained in the numbers of bursts detected by different surveys. For any one survey, the number of detected FRBs puts constraints on the rate of FRBs occurring above the detection threshold of that survey. This rate can be rescaled to a different survey with a different detection threshold and confronted with the observed number of bursts for that survey. However, the rescaling depends on the parameter $\alpha$ and thus the number of bursts detected by two or more surveys puts constraints on $\alpha$.

To include these constraints in our analysis we introduce the FRB rate explicitly as an unknown parameter. Since the rate observable by a given survey depends on various properties of the survey, we define the rate $r_0$ occurring above the detection threshold of a hypothetical survey described by a system temperature $T_{\mathrm{sys},0} = 1\,\mathrm{K}$, a gain $G_0 = 1\,\mathrm{K}\,\mathrm{Jy}^{-1}$, $n_{\mathrm{p},0}=2$ observed polarizations, a bandwidth $B_0 = 1\,\mathrm{MHz}$, and a signal-to-noise threshold $s_{\mathrm{min},0} = 1$. As explained by \citet{connor-2016b}, the FRB rate above the detection threshold of the $I$-th survey is then a rescaled version of this rate, namely
\begin{align}
	r_I &= r_0 \, \left(\frac{T_{\mathrm{sys},I}}{T_{\mathrm{sys},0}} \, \frac{G_0}{G_I} \, \sqrt{\frac{n_{\mathrm{p},0} \, B_0}{n_{\mathrm{p},I} \, B_I}} \, \frac{s_{\mathrm{min},I}}{s_{\mathrm{min},0}}\right)^{-\alpha}\nonumber\\
	&= r_0 \, \left(\frac{T_{\mathrm{sys},I}}{G_I} \, \sqrt{\frac{2\,\mathrm{MHz}}{n_{\mathrm{p},I} \, B_I}} \, s_{\mathrm{min},I} \, \mathrm{Jy}\right)^{-\alpha}.
	\label{eq:rescaling}
\end{align}

The expected number of FRBs detected by the $I$-th survey will then be
\begin{equation}
	M_I = r_I \, \Omega_I \, T_I,
\end{equation}
where $\Omega_I$ is the angular size of the survey's field of view and $T_I$ is the time spent surveying. The likelihood for the actual number of FRBs observed in this survey is then a Poissonian distribution with this expectation value,
\begin{equation}
	P(n_I|r_0,\alpha) = \frac{{M_I}^{n_I}}{n_I!} \, \mathrm{e}^{-M_I}.
\end{equation}
For $N$ surveys the complete likelihood again becomes a product of the likelihoods for the individual surveys,
\begin{equation}
	P(\vec{n}|r_0,\alpha) = \prod_{I=1}^N P(n_I|r_0,\alpha),
\end{equation}
and can be used to put constraints on the distribution of flux densities via the parameter $\alpha$, as well as on the overall rate of FRBs, here parameterized as the rate above the detection threshold of our hypothetical survey, $r_0$.

\subsection{Posterior}
\label{sec:posterior}

To get the complete set of constraints on the distribution of flux densities, both from the observed signal-to-noise ratios and from the detection numbers of different surveys, we combine the results of Sects.~\ref{sec:snr} and \ref{sec:number}. We write the joint likelihood for the number of observed FRBs and their signal-to-noise ratios as
\begin{align}
	\mathcal{P}(\vec{s},\vec{n}|r_0,\alpha) &= \mathcal{P}(\vec{s}|\vec{n},r_0,\alpha) \, P(\vec{n}|r_0,\alpha)\nonumber\\
	&= \mathcal{P}(\vec{s}|\vec{n},\alpha) \, P(\vec{n}|r_0,\alpha).
\end{align}
If we assume flat priors for $r_0$ and for $\alpha > 0$, this likelihood is proportional to the joint posterior for the parameter $\alpha$ and the rate $r_0$.

The likelihood for observed fluxes within a survey obviously only gives us constraints if the survey has in fact detected at least one FRB. Note, however, that we can in principle include surveys without FRB detection by setting
\begin{equation}
	\mathcal{P}(\vec{s}|n=0,s_\mathrm{min},\alpha) = 1
\end{equation}
and thus still use them to constrain the parameter $\alpha$ via their implications on the FRB rate above their detection thresholds. Similarly, the numbers of FRBs detected by different surveys only have implications for the parameter $\alpha$ if we assume that the surveys observe the same source population, described by the same rate $r_0$. This assumption will in general be violated if different surveys have different frequency coverage or different observational strategies. Specifically, the observations of a deep and narrow survey will in general not be described by the same statistics as those of a shallow and wide survey, as explained by \citet{connor-2016c}. Care is thus warranted when comparing detection numbers of qualitatively different surveys. Such an attempt will require more parameters or simply setting
\begin{equation}
	P(n|r_0,\alpha) = 1
\end{equation}
for all surveys that are not expected to be described by $r_0$.

\section{Data and results}
\label{sec:results}

We make use of 15 observed FRBs from seven surveys. For definiteness, we list all values used in our calculation in Tables~\ref{tab:surveys} and \ref{tab:FRBs}. For the likelihood of the numbers of detected FRBs, we only make use of two dedicated pulsar surveys with well-defined characteristics, namely the High Time Resolution Universe Pulsar Survey (HTRU; \citealt{keith-2010}) at the Parkes telescope and the Pulsar ALFA survey (PALFA; \citealt{cordes-2006}) at the Arecibo Observatory. We choose these two surveys because for most other discovered FRBs it is hard to estimate the surveying period $T$ that has been searched for FRBs, especially in the case of non-detections. The survey of \citet{masui-2015b} is similarly well-defined, but sensitive to different frequencies. We assume our parameter $r_0$ to describe the rate at frequencies around 1.4\,GHz and do not want to make any assumption about the relation between this rate and the rate at 800\,MHz, which is the central frequency of \citet{masui-2015b}.

Even for HTRU and PALFA, the parameters needed in Eq.~\eqref{eq:rescaling} are defined somewhat ambiguously. To avoid building complicated models of the telescopes and surveys, we generally opt for simple choices that can be made consistently for both surveys. Specifically, this means that we do not include any estimate of the sky temperature due to the Milky Way in the values we assume for the system temperature $T_\mathrm{sys}$. For the gain $G$ we use the arithmetic mean of the gains corresponding to the beam centres of the multibeam receivers. The bandwidth $B$ does not include frequencies deemed unusable by the surveying team and the angular size of the field of view $\Omega$ is intended to approximate the area within the half-maximum beam power. The resulting numerical values are listed in Table~\ref{tab:surveys}. Other reasonable choices for these parameters will typically lead to deviations on the order of 10\%. Note that the exact definition of each parameter does not impact the results as long as the same definition is used for all surveys that are being compared.

For the constraint on $\alpha$ coming from the likelihood for the observed signal-to-noise ratios, we can use all detected FRBs, as long as there is a well-defined signal-to-noise cutoff $s_\mathrm{min}$.
Since we are investigating the population of sources, we are not including repeated bursts from the same object \citep{spitler-2016}. We also exclude the single burst detections by \citet{lorimer-2007} and \citet{keane-2011}, since no definitive value of $s_\mathrm{min}$ can be determined. We list the values of $s$ and $s_\mathrm{min}$ that we use in Table~\ref{tab:FRBs}.

\begin{figure}
	\includegraphics{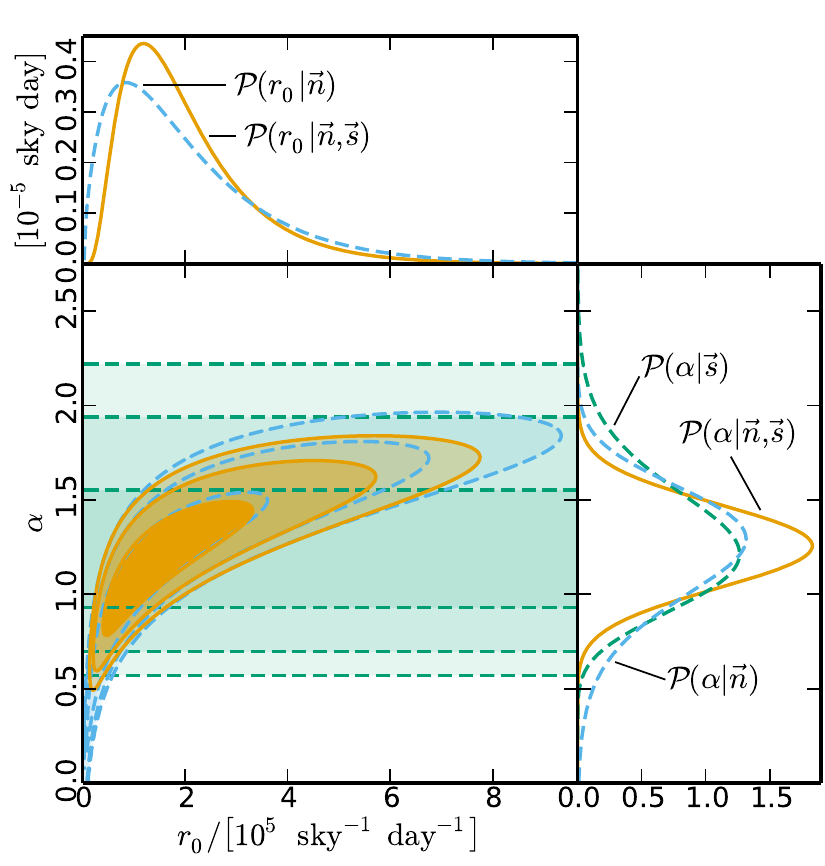}
	\caption{\label{fig:2dpost}Posterior distribution for the parameter $\alpha$ describing the distribution of FRB flux densities and the FRB rate $r_0$. The bottom left panel shows the two-dimensional posterior for both parameters, the smaller panels show the marginalised posteriors for each parameter individually. We show separate curves and contours for the constraints coming from the signal-to-noise ratios $\vec{s}$ (green dashed), from the detection numbers $\vec{n}$ (blue dashed), and their combination (orange solid). The contour lines show the 68\%, 95\%, and 99\% confidence regions.}
\end{figure}

After calculating the two-dimensional posterior for $\alpha$ and $r_0$, we derive the final constraint on $\alpha$ by marginalising over $r_0$ and vice versa. These posterior distributions are shown in Fig.~\ref{fig:2dpost}.

\begin{table*}
\begin{minipage}{116mm}
\begin{center}
	\caption{Parameters assumed for the FRB surveys. See Sect.~\ref{sec:number} for the meaning of the symbols.}
	\label{tab:surveys}
	\newdimen\digitwidth
	\setbox0=\hbox{\rm 0}
	\digitwidth=\wd0
	\catcode`*=\active
	\def*{\kern\digitwidth}
\begin{tabular}[tmb]{ccccccccc}
	\hline
	survey$^1$& $s_\mathrm{min}$& $T_\mathrm{sys}/\mathrm{K}$& $G/(\mathrm{K/Jy})$& $n_\mathrm{p}$& $B/\mathrm{MHz}$& $\Omega/(\mathrm{deg}^2)$& $T/\mathrm{h}$& $n$\cr
	\hline
	HTRU [1]& 10& 23& 0.64& 2& 340& $13\times0.043*$& 3650& 9\cr
	PALFA [2]& *7& 30& 8.5*& 2& 300& $*7\times0.0027$& *886& 1\cr
	\hline
\end{tabular}

\medskip
$^1$~[1]~\citealt{champion-2015,thornton-2013,keith-2010}; [2]~\citealt{spitler-2014,cordes-2006}
\end{center}
\end{minipage}
\end{table*}

\begin{table}
\begin{center}
	\caption{Parameters of each individual FRB used in our calculation. The signal-to-noise ratios $s$ are taken from the FRBcat website$^1$ \citep{petroff-2016}.}
	\label{tab:FRBs}
	\newdimen\digitwidth
	\setbox0=\hbox{\rm 0}
	\digitwidth=\wd0
	\catcode`*=\active
	\def*{\kern\digitwidth}
\begin{tabular}[tmb]{lccc}
	\hline
	name& $s$& $s_\mathrm{min}$& survey$^2$\cr
	\hline
	FRB090625& 30& 10& [1]\cr
	FRB110220& 49& 10& [1]\cr
	FRB110626& 11& 10& [1]\cr
	FRB110703& 16& 10& [1]\cr
	FRB120127& 11& 10& [1]\cr
	FRB121002& 16& 10& [1]\cr
	FRB130626& 21& 10& [1]\cr
	FRB130628& 29& 10& [1]\cr
	FRB130729& 14& 10& [1]\cr
	FRB121102& 14& *7& [2]\cr
	FRB010125& 17& *7& [3]\cr
	FRB131104& 30& *8& [4]\cr
	FRB140514& 16& 10& [5]\cr
	FRB150418& 39& 10& [6]\cr
	FRB110523& 42& *8& [7]\cr
	\hline
\end{tabular}

\medskip
$^1$~\url{http://www.astronomy.swin.edu.au/pulsar/frbcat/}\\
$^2$~[1]~\citealt{champion-2015,thornton-2013,keith-2010}; [2]~\citealt{spitler-2014,scholz-2016}; [3]~\citealt{burke-spolaor-2014}; [4]~\citealt{ravi-2015}; [5]~\citealt{petroff-2015}; [6]~\citealt{keane-2016}; [7]~\citealt{masui-2015b,connor-2016b}
\end{center}
\end{table}

\section{Discussion}
\label{sec:discussion}

Figure~\ref{fig:2dpost} shows a strong correlation between the FRB rate and the slope parameter $\alpha$. This can be understood in terms of regions of parameter space that are in tension with the data. If the rate of FRBs is high, a shallow flux density distribution will overpredict the number of FRBs occurring at high signal-to-noise ratios. If the rate is low, on the other hand, a steep distribution will underpredict the number of FRBs occurring above the detection threshold of current surveys, especially at high signal-to-noise values.

The figure also shows the posterior distributions that are obtained if only the detection numbers or the signal-to-noise ratios are used, instead of their combination. Obviously, the signal-to-noise ratios alone do not constrain the FRB rate at all. Thus, the corresponding contours appear as horizontal lines in the main panel of the figure. And even for the parameter $\alpha$, the main constraint comes from the comparison of the detection numbers for the two surveys HTRU and PALFA. Using the signal-to-noise ratios in addition does, however, add some information, in that it rules out close-to-flat flux distributions and very low rates.

The full posterior for the parameter $\alpha$ still allows a wide range of values. The 95\% confidence interval is
\begin{equation}
	0.8 \leq \alpha \leq 1.7.
\end{equation}
This is to be contrasted with recent results from the literature. \citet{caleb-2016}, for example, find $\alpha = 0.9 \pm 0.3$ and \citet{li-2016} claim $\alpha = 0.14 \pm 0.20$. We stress that our constraints are model-independent in the sense that we have not assumed any specific relation between flux density, burst duration, and dispersion measure. An important conclusion of our analysis is that the simplest possible model for the distribution of FRBs, constant density in Euclidean space, is consistent with current data. Of course this does not mean that it is proven to be correct, but it does mean that any extension to this model is not data-driven but has to be motivated independently. This finding is consistent with the qualitative conclusions of \citet{katz-2016a} and \citet{katz-2016b}.

As a byproduct of our attempt to constrain $\alpha$, we also obtain constraints on the rate of FRBs at 1.4\,GHz. The 95\% confidence interval for our parameter $r_0$ is
\begin{equation}
	4.8\times10^4\,\mathrm{sky}^{-1}\,\mathrm{day}^{-1} \leq r_0 \leq 5.3\times10^5\,\mathrm{sky}^{-1}\,\mathrm{day}^{-1}.
\end{equation}

It may be worth noting that the constraints on the FRB rate tighten up somewhat if the parameter $\alpha$ is fixed. In the constant-density Euclidean model, for example, the 95\% confidence limit on the rate is
\begin{equation}
	1.6\times10^5\,\mathrm{sky}^{-1}\,\mathrm{day}^{-1} \leq r_{0,\alpha=3/2} \leq 5.4\times10^5\,\mathrm{sky}^{-1}\,\mathrm{day}^{-1}.
\end{equation}
For any specific survey this rate has to be rescaled according to Eq.~\eqref{eq:rescaling}. As an example we calculate the FRB rate above the detection threshold of the HTRU survey, again for $\alpha=3/2$,
\begin{equation}
	1.9\times10^3\,\mathrm{sky}^{-1}\,\mathrm{day}^{-1} \leq r_{\mathrm{HTRU},\alpha=3/2} \leq 6.3\times10^3\,\mathrm{sky}^{-1}\,\mathrm{day}^{-1}.
\end{equation}
These are slightly lower values than the range derived by \citet{champion-2015}. We stress again that all rates we calculate are subject to a signal-to-noise cutoff. Converting them to rates above a given fluence is impossible without making further assumptions.

\section*{Acknowledgements}

We thank Vikram Ravi and Kiyo Masui for helpful discussions and Jonathan Katz, Evan Keane, and an anonymous referee for useful comments on the manuscript. This research has made use of NASA's Astrophysics Data System. The figure was produced using the \texttt{matplotlib} library \citep{hunter-2007} and we acknowledge use of the \textsc{FRBcat} database \citep{petroff-2016}. We acknowledge NSERC support.

\bibliographystyle{myaa}
\bibliography{alpha}

\bsp
\label{lastpage}
\end{document}